\documentclass[%
 reprint,
 amsmath,amssymb,
prl,
]{revtex4-2}

\usepackage{lipsum}
\usepackage{graphicx}
\usepackage{bm}
\usepackage{color}
\usepackage{times}
\usepackage{amsmath}
\usepackage{float}
\usepackage{hyperref}
\usepackage{ulem} 
\usepackage[dvipsnames]{xcolor}
\usepackage{hyperref}

\begin{document}


\title{Experimental observation of flow instability control by metamaterial subsurfaces}

\author{Melanie Keogh$^1$}
\author{Jensen McTighe$^2$}
\author{Jason Dahl$^2$}
\author{Osama R. Bilal$^1$}
\email{osama.bilal@uconn.edu}
\affiliation{$^1$School of Mechanical, Aerospace, and Manufacturing Engineering, University of Connecticut, Storrs , CT , 06269, USA.}
\affiliation{$^2$Department of Ocean Engineering, University of Rhode \\Island, Narragansett, 02881, RI, USA}

\date{\today}

\begin{abstract}
Flow instabilities within a fluid flow can cause laminar-to-turbulent transition over surfaces. These instabilities can result from upstream, wake-generating disturbances, leading to increased drag and turbulence-induced energy losses. Flow control strategies can address these issues through active methods, requiring energy input, or passive systems, which operate without added input. Here, we present a passive approach to flow control using embedded phononic metamaterials to alter vortex instability development, without changing the outer-surface's texture, roughness or compliance. Experiments confirm that our subsurface can suppress vortex growth at target frequencies, demonstrating the potential for energy-efficient flow management with phononic subsurfaces.
\end{abstract}

\maketitle

\begin{figure*}
\includegraphics{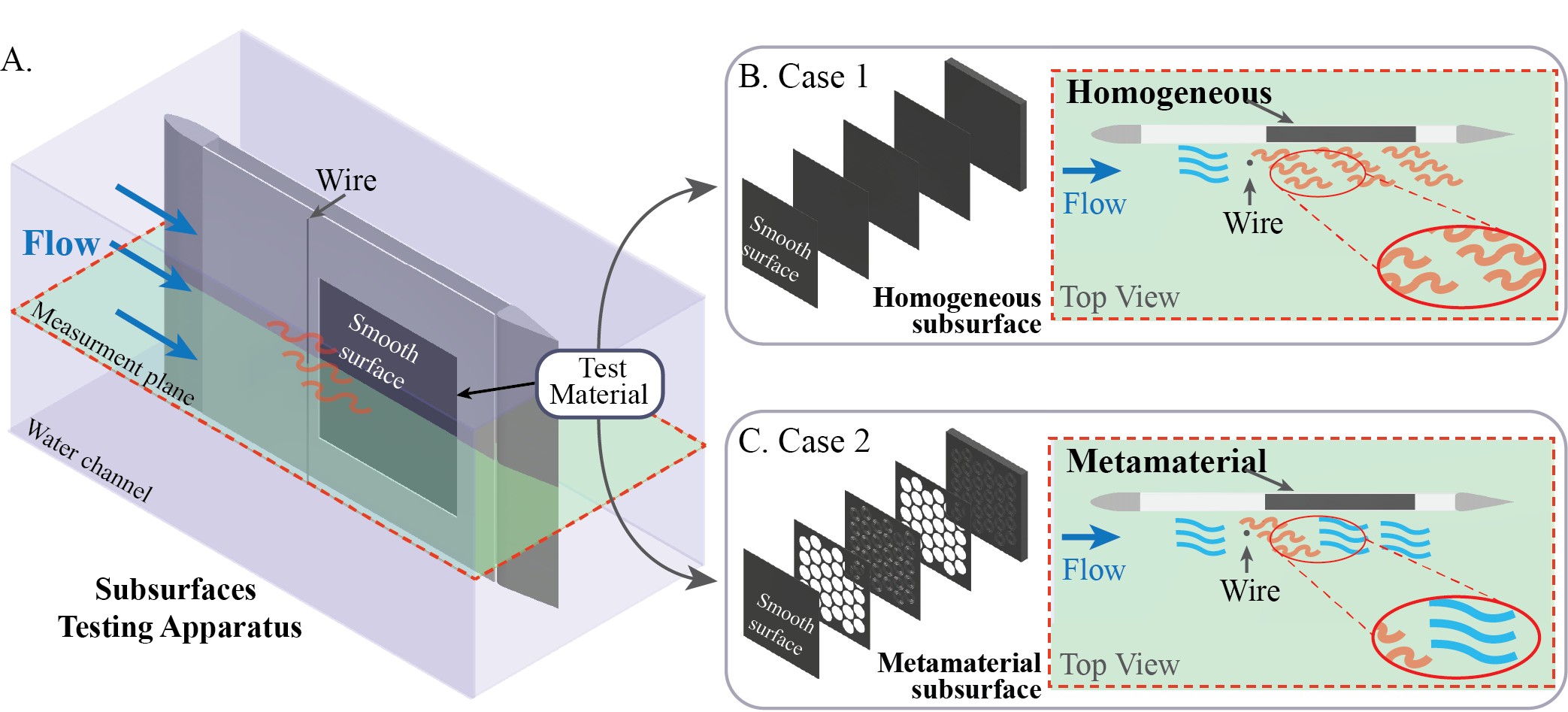}
\caption{\label{fig:concept}\textbf{Concept.} (A) Schematic of the phononic subsurface testing apparatus. Laminar flow enters the channel from the left and encounters a thin wire that disrupts the flow such that vortices are generated in the flow over the test material with a smooth outer layer and various subsurfaces (cases 1 and 2). Velocity field measurements are conducted in a cross-sectional plane perpendicular to the subsurface test material. (B) Exploded view of homogeneous subsurface layers and conceptual schematic illustrating how the homogeneous subsurface has no measurable effect on the flow field within the measurement plane. (C) Exploded view of metamaterial subsurface layers and conceptual schematic illustrating how the metamaterial subsurface interacts with the flow field such that targeted vortex shedding frequencies are attenuated.}
\end{figure*}

The formation and growth of flow instabilities on submerged surfaces and in long-range pipes can lead to significant drag forces as a result of turbulence-induced energy losses. Flow control focuses on inducing a desirable change in the flow field by active or passive means and can be characterized by the flow receptivity, or the degree to which the flow, particularly a boundary layer, is susceptible to external disturbances \cite{arnal2000laminar, hussein2015flow, thomas2022optimizing}. Such disturbances include Tollmien–Schlichting (TS) waves, which may arise within a bounded shear flow \cite{hogberg1998secondary, hussein2015flow,  michelis2023interaction}, and cross-flow vortices \cite{hogberg1998secondary, thomas2022optimizing}, which can form due to separation or upstream disturbances.  Active methods, such as pumps, actuators, and vortex generators, are widely used to promote favorable flow manipulation through momentum injection and alteration of surface pressures such that desirable flow characteristics or hydrodynamic forcing is achieved. However, due to the active nature of these methods, energy input is required
\cite{yan2011water, gad2019coherent, wang2020tunable, fan2020reinforcement,  liu2020active,  hajipour2022active, mao2022active, vivas2022fluidic, thomas2022optimizing, lin2024active, subedi2024turbulent}. In contrast, a passive approach to flow control enables flow manipulation without spending additional energy \cite{sun2014decay, bocanegra2018engineered, che2019control, zhu2020flow, zaresharif2021cavitation, santos2021passive, fatahian2023aerodynamic, willey2023multi, michelis2023attenuation, kianfar2023local, zani2024turbulent}. While this passive approach is desirable from an energy perspective, many existing flow control methodologies rely on changing the compliance, texture, or roughness of surfaces directly in contact with the flow \cite{kandlikar2005characterization, zhou2012effects, huebsch2012dynamic,  bixler2013shark, casadei2014harnessing, terwagne2014smart, afroz2016experimental, alves2020suppression, shi2024dynamic, jarmon2024wind}, where these surface manipulations can prove challenging to scale up and maintain in application. 

An alternative approach to flow control relies on engineering the dynamics of the \textit{subsurface} of the moving object, in contrast to its \textit{outer surface}. By designing a solid with a specific elastic response, one can utilize the interaction of the fluid and the structure to enact a desirable response in the fluid by utilizing subsurface phonons. Such an approach capitalizes on using phononic media where, for example, the solid material may delay or promote the laminar-to-turbulent transition within a fluid flow. Phononic subsurface designs for flow control are mostly based on a phononic crystal (i.e., Bragg-scattering band gaps \cite{hussein2015flow, bilal2015design,barnes2021initial, willey2023multi, machado2024fluid, navarro2024flow} imposing a strict constraint on the size of the subsurface relative to the target wavelength.  Locally resonant phononic subsurfaces \cite{ kianfar2023phononic, kianfar2023local} can break the correlation between size and wavelength, providing subwavelength flow control and resulting in a smaller footprint. Although many numerical simulations have been made to theorize the behavior of such surfaces \cite{hussein2015flow, bilal2015design,barnes2021initial, kianfar2023phononic,willey2023multi, kianfar2023local,machado2024fluid, navarro2024flow}, the use of a smooth passive phononic subsurface to control a flow field has not been experimentally realized. Here, we present a methodology for designing an ultra-thin locally resonant metamaterial subsurface and show experimentally that it can passively interact with a fluid flow and attenuate vortex formation at targeted frequencies.

Phononic media are artificial structures composed of building blocks, or unit cells, that repeat in space with a frequency-dependent dynamical response. A key characteristic of phononic media is the existence of band gaps within their frequency spectrum. An excited frequency within a band gap is attenuated by the phononic material. In contrast to a homogeneous material, phononic media can influence the stability of flow, as a subsurface, through constructive and destructive interferences \cite{ hussein2015flow, alves2020suppression, barnes2021initial, shi2024dynamic, machado2024fluid}. A phononic subsurface is a phononic crystal \cite{sigalas1993band, kushwaha1993acoustic} or an acoustic / elastic metamaterial \cite{liu2000locally,rupin2014experimental} encased in a smooth, homogeneous outer surface (Fig.\ref{fig:concept}). Such phononic subsurfaces can be designed to interact with a fluid flow across its outer surface without the need to manipulate the texture or geometry of the surface interacting directly with the flow. For example, vortex formation at a specific frequency can be suppressed using a phononic subsurface compared to a homogeneous subsurface despite both structures having the same smooth outer surface (Fig.\ref{fig:concept}B-C).

\begin{figure*}
\includegraphics{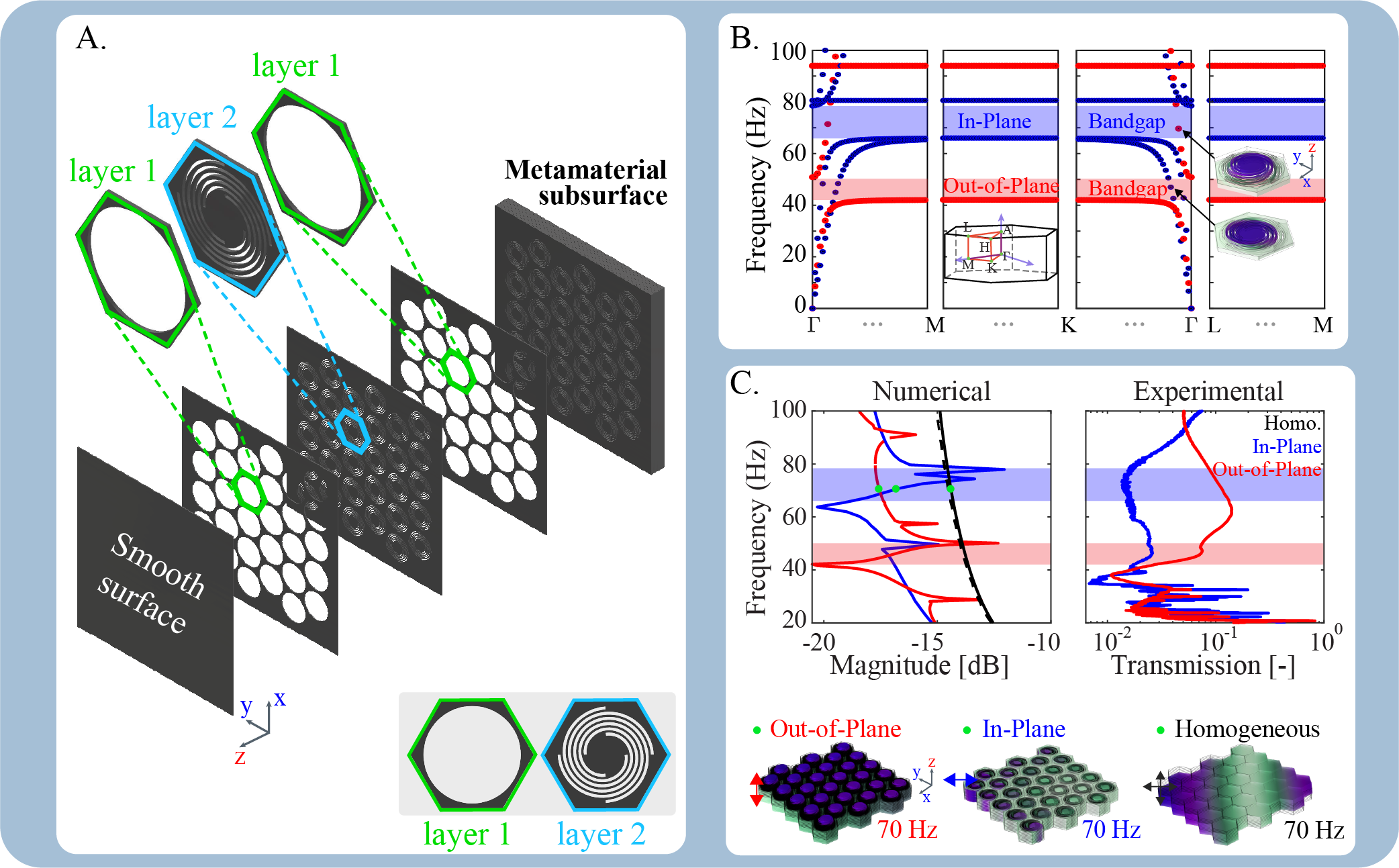}
\caption{\label{fig:2}\textbf{Metamaterial Subsurface Phononic Characterization:}  (A) Exploded view of metamaterial subsurface highlighting the layer geometries. The metamaterial subsurface is composed of alternating layers of hexagonally packed holes and spirals.  (B) Dispersion curves for the double layer unit cell considering periodicity in all directions. In-plane and out-of-plane modes and band gaps are highlighted in blue and red, respectively. The inset shows mode shapes at 47 and 70 Hz.  (C) Numerical frequency response function (left) and experimental transmission (right). Homogeneous frequency responses for out-of-plane and in-plane excitations are shown in solid and dotted black lines, respectively. Metamaterial frequency responses and band gaps for in-plane excitations are highlighted in blue. Metamaterial frequency responses and band gaps for out-of-plane excitations are highlighted in red. Mode shapes of finite structures are shown. First and second mode shapes show the metamaterial structure excited at 70 Hz out-of-plane and in-plane, respectively. Third mode shape shows homogeneous finite structure excited in both out-of-plane and out-of-plane at 70 Hz.}
\end{figure*}

\begin{figure*}
\includegraphics{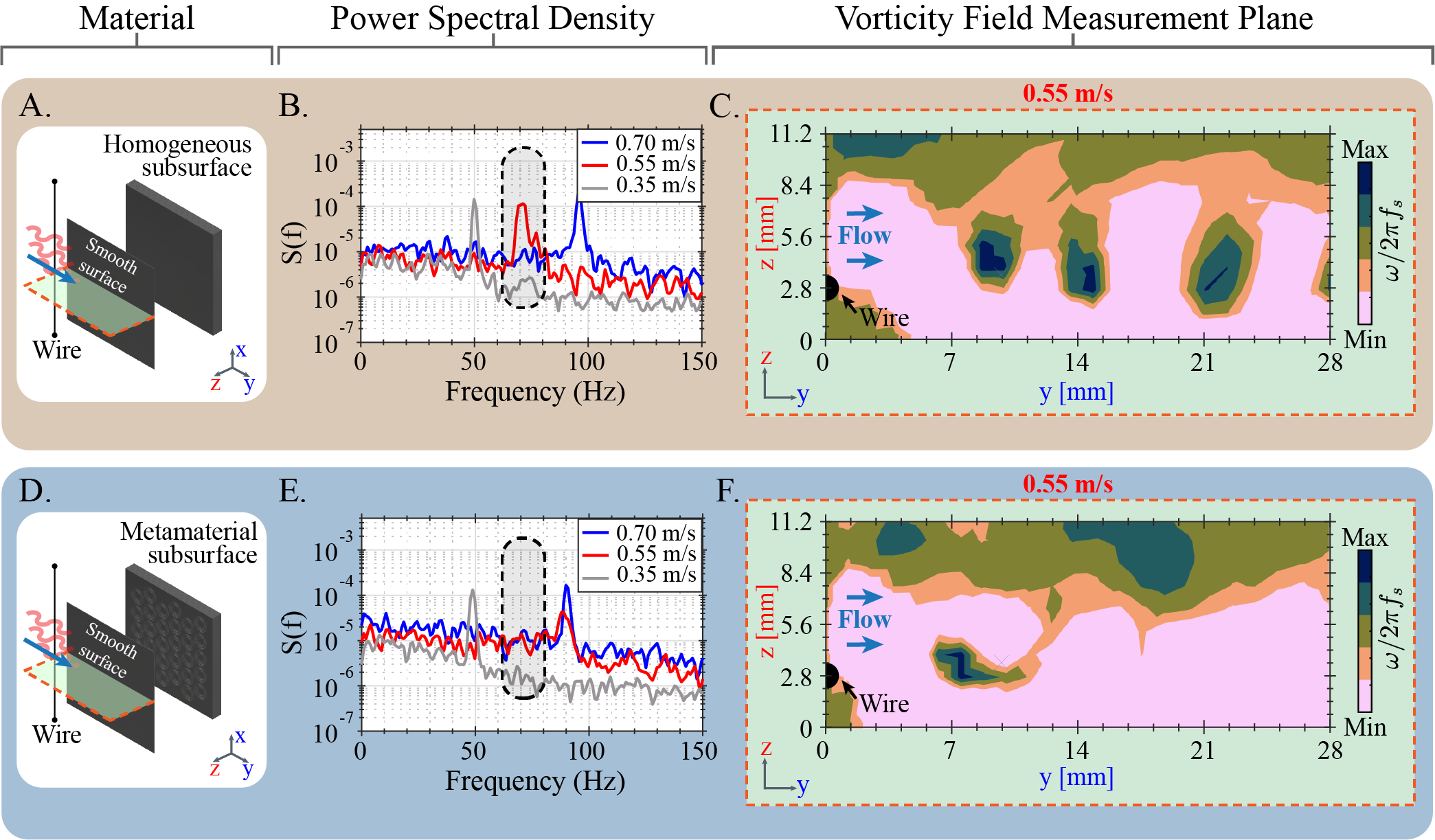}
\caption{\label{fig:3}\textbf{Metamaterial Subsurface Flow Experiment:}  (A) Schematic of homogeneous subsurface and disturbance generated by flow over wire. (B) Power spectral density computed from the time history of the velocity component in the $z$ direction measured at 9.6 mm (7 wire diameters) downstream of the wire center and the wire located 2 diameters from the homogeneous test material in flow velocities of .35 m/s, .55 m/s, and .70 m/s shown in gray, red, and blue, respectively. (C) Resulting vorticity field near the homogeneous subsurface material with .55 m/s flow velocity. Regular vortex formation corresponding to the Strouhal frequency at this flow speed is visible in the wake. (D) Schematic of metamaterial subsurface and disturbance generated by flow over wire. (E) Power spectral density computed from the time history of the velocity component in the $z$ direction measured at 7 wire diameters downstream of the wire center and the wire located 2 diameters from the metamaterial exterior surface in flow velocities of .35 m/s, .55 m/s, and .70 m/s shown in gray, red, and blue, respectively. (F) Resulting vorticity field near the metamaterial with .55 m/s flow velocity. A clear reduction in the visible presence of vortices in the wake of the wire is shown.}
\end{figure*}

We begin our analysis by considering the dynamics of a single metamaterial unit cell. The metamaterial consists of layers of hexagonally packed Archimedean spirals \cite{bilal2017bistable,bilal2017reprogrammable, foehr2018spiral,bilal2020enhancement, bilal2020flexible, jiang2020randomized, krushynska2021dissipative,  tian2021merging, li2021topological, jin2022deep, gu2023lightweight, kheybari2023harnessing, jiang2024spatial} separated by layers of hexagonal circular holes. The circular holes above and below each spiral unit cell act as void regions, allowing the spiraling cores to vibrate freely. Both the spiral layers and the hole layers have a thickness of 1.5 mm, each. The metamaterial array is sandwiched between two homogeneous smooth outer surfaces (Fig.\ref{fig:2}A). We assume Bloch periodic boundary conditions in all three directions of the unit cell in the calculations of the dispersion curves. We further sort individual modes into in-plane (blue dots) and out-of-plane (red) based on their dominant displacement direction (Fig.\ref{fig:2}B). We note the existence of polarization-dependent band gaps within our target frequency range below 100 Hz. An out-of-plane band gap exists from 42 to 51 Hz and an in-plane band gap from 66 to 78 Hz for all considered wavenumbers. To help visualize these polarization band gaps, we plot the unit cell mode shapes at two frequencies: (1) 47 Hz, a frequency within the out-of-plane band gap exhibiting in-plane motion, and (2) 70 Hz, a frequency within the in-plane band gap exhibiting out-of-plane motion.

In order to validate our infinite unit-cell dispersion analysis, we numerically simulate the dynamic response of a finite structure composed of 6x5x9 unit cells in the \textit{x-y-z} directions, respectively.  We excite the finite metamaterial array in both the in-plane and out-of-plane directions. For the in-plane case, we excite the finite structure in \textit{x-y} directions at the corner of the top surface and measure the transmission at the opposite corner (Fig.\ref{fig:2}C-left).  For the out-of-plane case, we excite the finite structure in the \textit{z} direction at the top center of the structure and measure the transmission at the bottom center of the structure (Fig.\ref{fig:2}C-left). The displacement amplitudes for both in-plane and out-of-plane cases show a clear dip (i.e., antiresonance) at the lower edge of their respective band gaps in good agreement with the dispersion relation (Fig.\ref{fig:2}C-left). We also observe a resonance peak at the upper edge of the predicted band gap for both cases.  For comparison, we repeat the same simulation for a homogeneous block of material composed of the same base material as our metamaterial. The resulting transmission amplitudes show no clear resonance or antiresonance at the frequency of interest (Fig.\ref{fig:2}C-left). In addition, we plot the mode shapes of both finite structures (i.e., metamaterial and homogeneous)  when excited at the target frequency of 70 Hz (Fig.\ref{fig:2}C). For the in-plane excitation, the wave attenuates after a few unit cells in the \textit{x-y} plane, while propagating in the \textit{z} direction at the excitation point. More importantly, the excitation frequency of 70 Hz is a pass band frequency for only the out-of-plane polarization, resulting in displacement amplification in the out-of-plane direction. In contrast, both in-plane and out-of-plane excitations of the homogeneous sample result in wave propagation at the excited frequency. These mode shapes show that the metamaterial, when excited out-of-plane at the target frequency on its surface, will generate an amplified out-of-plane surface response.

In order to validate our numerical models, both infinite and  finite, we test the proposed design experimentally. We  fabricate our metamaterial sample by assembling alternating hexagonally packed spiral layers and hole layers. We use epoxy resin as an adhesive between each layer. Each layer consists of a 5 x 6 array of a repeated geometric feature surrounded by a homogeneous border with a layer thickness of 1.5 mm. We add homogeneous layers to the top and the bottom of the sample. The finite metamaterial sample consists of 21 layers (9 layers with spiral patterns, 10 layers of holes, and two homogeneous layers on top and bottom) with a total thickness of 31.5 mm. It is worth noting that the thickness of our metamaterial is 2 orders of magnitude thinner than the originally proposed phononic subsurface materials \cite{hussein2015flow}, while operating at approximately 2 orders of magnitude lower frequency (approximately 4 orders of magnitude improvement with the appropriate scaling). We also bond 21 homogeneous layers composed of the same material using epoxy resin as a reference homogeneous sample.  We excite both the metamaterial array and the homogeneous sample at the center of one side and measure the displacement of the perpendicular and opposite sides to the excitation, for both out-of-plane and in-plane cases, respectively. For these measurements, we use a scanning laser Doppler vibrometer. In general, the measurements match well with both band gap predictions (Fig.\ref{fig:2}C-right). We note a small shift in the out-of-plane band gap top edge, most likely due to the epoxy adhesive. It is also worth noting that the out-of-plane band gap region is also captured in the in-plane measurement due to the excitation polarization. 

In order to demonstrate the utility of the metamaterial in interacting with a flow in a favorable manner, we conduct multiple experiments consisting of a test plate mounted in a recirculating flow channel as shown in Figure \ref{fig:concept}. Laser Doppler velocimetry (LDV) measurements confirm a uniform vertical and horizontal flow profile within the flow channel test section with background turbulence intensities less than 2\% of the mean free stream and very low background energy content in the flow field at expected excitation frequencies in the experiments.  The same two homogeneous and metamaterial samples that were tested previously in dry conditions are used in the flow channel experiments. It is important to note that, both samples are manufactured following identical procedures,  materials,  outer dimensions, and more importantly, the same surface roughness. Each sample is fixed in a streamlined frame mounted in the middle of the flow channel such that the outermost surface of the test sample is in direct contact with the incoming flow (Fig.\ref{fig:concept}). A thin wire with a radius of 0.7 mm is positioned over the upstream edge of the sample and 4 times the wire radius from the sample surface.  As the fluid flows over the wire, vortex shedding occurs at a frequency proportional to the flow speed. We consider three different velocities of the fluid flow excitation for each case. Flow speeds of 0.35 m/s, 0.55 m/s, and 0.70 m/s correspond to expected vortex shedding frequencies of 50 Hz, 79 Hz, and 100 Hz, respectively, assuming a constant Strouhal number of 0.2.  The wire was tensioned such that the fundamental natural frequency of the wire was over twice the highest expected shedding frequency to avoid lock-in effects of the wire's natural frequencies and vortex shedding. The wire was also positioned within the developed boundary layer, such that the flow speed encountered by the wire was reduced. Therefore, the measured vortex shedding frequency was expected to be slightly less than the expected frequency based on the assumed Strouhal number. To measure the flow field, we obtain time-resolved particle image velocimetry (PIV) measurements of the velocity field in the wake of the wire and at the surface of the test material.  From the velocity field measurements, we compute the power spectrum of the velocity component in the $z$ direction (Fig.\ref{fig:3}B\&E) and the instantaneous vorticity field (Fig.\ref{fig:3}C\&F). In the reference case of the homogeneous subsurface, flow velocities of .35 m/s, .55 m/s, and .70 m/s correspond to measured vortex shedding frequencies of 50 Hz, 70 Hz, and 96 Hz respectively. In the metamaterial case, for the two flow speeds corresponding to an in-plane pass band frequency (0.35 m/s, and 0.70 m/s, with measured peak frequencies of 49 and 90 Hz, respectively) the shedding behavior remained relatively unchanged compared to the homogeneous case. However, for the flow speed of 0.55 m/s, which corresponds to a measured frequency of 70 Hz, the power spectral density shows a clear peak with strong frequency content for the homogeneous sample and attenuation of the the power spectrum for the metamaterial array at the same frequency, as seen in the shaded regions of Figure \ref{fig:3}B\&E. Additionally, for the flow speed of 0.55 m/s in the metamaterial case, we observe a small peak around 88 Hz at a lower amplitude, implying that the shedding frequency has shifted to a larger value.  This potentially indicates a fluid-structure interaction between the metamaterial and the fluid, where the vortex shedding frequency has shifted to a higher frequency in the pass band of the metamaterial.  

To further elucidate how the flow field is impacted by the presence of the metamaterial subsurface, we show a snapshot in time of the vorticity field for both the homogeneous and metamaterial cases at a flow velocity of 0.55 m/s. In the homogeneous case, as the fluid flows over the wire, the upper half of a typical K\'arm\'an vortex street is visible with the positively signed vortices propagating downstream over time. However, in the metamaterial case, as the flow crosses the wire, only a single vortex visibly forms in the wake of the cylinder, briefly propagating downstream before quickly dissipating.  This vortex reappears periodically, but rapidly  dissipates downstream.  This behavior is consistent with our dispersion calculations and  frequency response function results. The excited mode by the flow at 70 Hz corresponds to an in-plane band gap and an out-of-plane mode that destructively interferes with vortex development within the flow near the surface of the metamaterial. Additionally, we note that since the wire is positioned at the edge of the test material, there are several centimeters of homogeneous material before the subsurface unit cells begin in the metamaterial.  As a result, we observe vortices forming upstream of where the unit cells begin, which then dissipate as they encounter the first unit cell.

In conclusion, we present an ultra-thin phononic  metamaterial subsurface and demonstrate its potential to destructively interact with a flow field such that vortex development and propagation is attenuated for target frequencies. Our results provide the first experimental evidence of the utilization of a phononic metamaterial subsurface to control vortex development in a fluid flow. Our design approach could be sought after in applications that require significant reduction in drag forces or the manipulation of object wakes. 


\end{document}